\documentclass[aps, twocolumn, prl, superscriptaddress]{revtex4-2}
\usepackage{amsmath}
\usepackage{mathtools}
\usepackage{physics}
\usepackage{datetime}
\newdate{date}{31}{05}{2023}
\usepackage{enumerate}
\usepackage{graphicx} 
\usepackage[hidelinks]{hyperref} 
\usepackage{xcolor}

\newcommand{\unit}[1]{\,\mbox{#1}}

\newcommand{\kHz}{\unit{kHz}}
\newcommand{\MHz}{\unit{MHz}}

\newcommand{\uK}{\unit{$\mu$K}}

\newcommand{\us}{\unit{$\mu$s}}

\newcommand{\degree}{\mbox{$^{\circ}$}}

\newcommand{\gnd}{$^1\mathrm{S}_0$ }
\newcommand{\exc}{$^3\mathrm{P}_1$ }

\newlength{\onecolfig}
\newlength{\twocolfig}
\makeatletter
\def\maketitle{
\@author@finish
\title@column\titleblock@produce
\suppressfloats[t]}
\makeatother

\begin{document}

\title{Many-body gap protection of motional dephasing of an optical clock transition}

\author{Zhijing Niu}
\affiliation{JILA, NIST, and Department of Physics, University of Colorado, Boulder, CO, USA}
\author{Vera M. Sch\"afer}
\affiliation{JILA, NIST, and Department of Physics, University of Colorado, Boulder, CO, USA}
\affiliation{Max-Planck-Institut f\"ur Kernphysik, Saupfercheckweg 1, 69117 Heidelberg, Germany}
\author{Haoqing Zhang}
\affiliation{JILA, NIST, and Department of Physics, University of Colorado, Boulder, CO, USA}
\affiliation{Center for Theory of Quantum Matter, University of Colorado, Boulder, CO, USA}
\author{Cameron Wagner}
\affiliation{JILA, NIST, and Department of Physics, University of Colorado, Boulder, CO, USA}
\author{Nathan R. Taylor}
\affiliation{JILA, NIST, and Department of Physics, University of Colorado, Boulder, CO, USA}
\author{Dylan J. Young}
\affiliation{JILA, NIST, and Department of Physics, University of Colorado, Boulder, CO, USA}
\author{Eric Yilun Song}
\affiliation{JILA, NIST, and Department of Physics, University of Colorado, Boulder, CO, USA}
\author{Anjun Chu}
\affiliation{JILA, NIST, and Department of Physics, University of Colorado, Boulder, CO, USA}
\affiliation{Center for Theory of Quantum Matter, University of Colorado, Boulder, CO, USA}

\author{Ana Maria Rey}
\affiliation{JILA, NIST, and Department of Physics, University of Colorado, Boulder, CO, USA}
\affiliation{Center for Theory of Quantum Matter, University of Colorado, Boulder, CO, USA}
\author{James K. Thompson}
\affiliation{JILA, NIST, and Department of Physics, University of Colorado, Boulder, CO, USA}

\usdate
\date{\today}

\begin{abstract}

Quantum simulation and metrology with atoms, ions, and molecules often rely on using light fields to manipulate their internal states. The absorbed momentum from the light fields can induce spin-orbit coupling and associated motional-induced (Doppler) dephasing, which may limit the coherence time available for metrology and simulation. We experimentally demonstrate the suppression of Doppler dephasing on a strontium optical clock transition by enabling atomic interactions through a shared mode in a high-finesse optical ring cavity. The interactions create a many-body energy gap that increases with atom number, suppressing motional dephasing when it surpasses the dephasing energy scale. This collective approach offers an alternative to traditional methods, like Lamb-Dicke confinement or Mössbauer spectroscopy, for advancing optical quantum sensors and simulations.

\end{abstract}

\maketitle
\vskip 0.5cm

In many applications for quantum simulation and metrology, it would be ideal to work in pristine conditions where the motional and internal degrees of freedom of atoms, ions or molecules are totally decoupled from each other.  For example, this would be an perfect scenario for atomic clocks or for the exploration of quantum magnetism. However, when dealing with optical transitions, photons impart significant momentum when used to manipulate the internal state of the atoms.  This so-called spin-orbit coupling, or coupling between motion and spin \cite{lin2011spin, kolkowitz2017spin}, can be a resource when under careful control, but it is just often a drawback that leads to single particle dephasing that limits the precision and fidelity of both quantum metrology and simulation experiments. 

Suppression of spin-orbit coupling or Doppler dephasing in fact has been a major driver of experimental efforts in the atomic physics community, leading to the development of laser-cooling techniques. Inspired by the pioneering work of Dicke \cite{Dicke1953} and Mössbauer \cite{mossbauer1959z,pound1960apparent}, experiments now use strong trapping potentials to spatially localize the atoms to much less than the wavelength of light they absorb and emit, see Fig.~\ref{fig1}a to c. 

However, strong traps can introduce additional decoherence mechanisms.  For instance, optical traps as used for neutral atoms can cause AC Stark shifts and light scattering \cite{campbell2017clock}, both of which must be controlled not only in quantum metrology, but also in state-of-the-art quantum computing and simulation settings with atoms, ions, and molecules. As such, it is extremely desirable to discover new and complementary ways to suppress motional dephasing. 

In this work, we experimentally demonstrate a novel collective mechanism in which an ensemble of strontium atoms interact via a shared optical mode of a ring cavity leading to the suppression of motional dephasing, see Fig.~\ref{fig1}b and d.  The suppression arises from a self-generated many-body energy gap that creates an energy penalty for evolving to states of lower symmetry \cite{Rey2008}.  
Such suppression was first observed recently in the context of a Bragg matter-wave interferometer, where dressing lasers were used to generate cavity-mediated momentum-exchange interactions that extended the coherence time between atomic ground momentum states separated in energy by only $\sim500$~kHz  \cite{luo_momentum-exchange_2024}.  Here we show that cavity-mediated exchange interactions can suppress Doppler dephasing for an optical clock transition with states separated by $400$~THz, see Fig.~\ref{fig1}f, thus providing a new path for quantum sensing using optical transitions \cite{SafranovaSearch2018}.

This new capability is part of the broader goal of understanding how to use many-body interactions to enhance quantum metrology and simulation. While the use of a many-body gap to prolong coherence between internal states has been explored in various settings including in standing wave optical cavities with strontium and rubidium atoms under tight confinement \cite{norcia_2018_science_jkt,davis_2020_prl_mss, young_2024_nature_jkt}, Coulomb interacting ions  \cite{franke2023quantum}, and atomic collisions \cite{AllredSERFMagnetometer2002, DeutschSpinSelf2010,smale2019observation,Huang2023,Huang2024}, to our knowledge this is the first time it has been observed using an optical transition without single-particle Lamb-Dicke suppression of motional dephasing.  This is achieved here by  the use of a ring cavity with position-independent atom-cavity coupling.

We also demonstrate the gap protection  mechanism  in a ring cavity beyond a two-level system by coupling to multiple optical excited states. The Hamiltonian of this extended system connects to models of modified neutrino oscillations in extreme astrophysical environments \cite{balantekin2023quantum,siwach2023entanglement}, relativistic quantum mechanical systems via spin-orbital coupling \cite{LeBlanc2013} or   condensed matter systems featuring a Dirac cone dispersion  such as  graphene \cite{Novoselov2011,Huang2016},  and topological insulators \cite{Hasan2010,Goldman2014} in a regime where interactions are dominant.

\begin{figure*}[!ht]
    \centering    \includegraphics[width=\twocolfig]{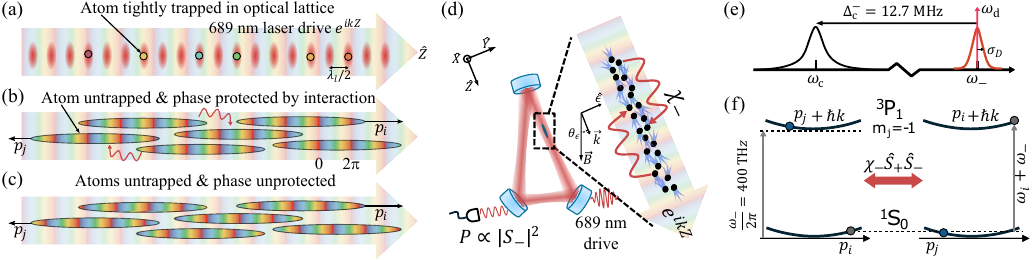}
    \caption{Experimental setup. (a) Lamb-Dicke suppression of motional dephasing.  An array of strontium atoms (dots) is tightly trapped in an optical lattice (red ovals) with wavelength $\lambda_l$  and placed in a superposition of a ground $^1$S$_0$ and optically excited state $^3$P$_1, m_j = -1$ by a laser field along $\hat{Z}$ whose spatial phase $e^{i k Z}$ is indicated via the color.  Since the atoms are localized, the motional state is unchanged, and the superposition state of atom $i$ at position $Z_i$ simply acquires the local phase  $e^{i k Z_i}$ indicated by the color of the dot aligning to the spatial phase of the excitation laser. (b) If the atoms are untrapped, we can model each atom as being a delocalized wave packet of approximately well-defined initial momenta $p_i$, $p_j$, etc.  The laser excitation now imprints a spatially varying phase on the superposition of the ground and excited states. (c)  Differences in kinetic energies cause the phases of different atoms to become randomized relative to each other and thus lead to a loss of collective coherence.  Allowing the atoms to exchange photons (red squiggles in (b)) suppresses the motional dephasing. (d) The exchange of photons is mediated by a ring cavity.  The laser drive at 689~nm excites the clock-wise mode of the cavity with linear polarization $\hat{\epsilon}$ and mode function $e^{i k Z}$.  A magnetic field $\vec{B}$ defines the quantization axis. We detect the optical power $P$ from the clockwise mode after the laser excitation to measure the collective atomic coherence $|S_-|^2$ versus time. (e) The bare cavity $\omega_c$ is detuned by $\Delta_c^-$ from the atomic transition frequency $\omega_-$ that is imhomogeneously broadened by $\sigma_D$ by the atomic motion (red). (f) Atoms can virtually exchange photons via the cavity, leading to an all-to-all effective spin-exchange interaction $\chi_-\hat{S}_+\hat{S}_-$ in which atoms change both their internal and momentum states.} 
    \label{fig1}
\end{figure*}

In our experiment, $N=10^6$ $^{88}\mathrm{Sr}$ atoms are laser cooled and trapped in a high-finesse optical ring cavity using a 1D optical lattice at $\lambda_l = 813$~nm, and  with radial temperature of $10\uK$, see Fig.~\ref{fig1}d and \cite{cline2022continuous}. The lattice serves to position the atoms within the cavity mode but will be switched off during the subsequent experiments so that the atoms can freely move inside the cavity mode and experience Doppler dephasing.

The cavity's resonance frequency $\omega_c$ is red-detuned from the  $\gamma=2\pi \times 7.5$~kHz linewidth  transition \exc - \gnd at frequency $\omega_0$ by $\Delta_c^0 = \omega_c - \omega_0 = -2\pi \times 14$ \MHz. The three-mirror cavity has a mode waist size of approximately $83~\mu$m, finesse $\mathcal{F}=6.2\times 10^3$, and  full-width-half-maximum (FWHM) linewidth is $\kappa = 2 \pi \times 266 \kHz$ at the transition wavelength of $\lambda=689$~nm for the relevant p-polarization mode $\hat{\epsilon}$ lying in the plane of the ring cavity, see Fig.~\ref{fig1}d. There is a vertically oriented magnetic field that sets the quantization axis and induces Zeeman splittings between the adjacent excited states $\ket{^3\mathrm{P}_1, m_j}$ levels by $\delta_B= 2\pi \times 1.3$~MHz.   Under this field, we define detunings of the cavity relative to each atomic transition frequency $\Delta_c^{-,0,+} = \omega_c -\left(\omega_0 + m_j \delta_B\right)$. The orientation of the magnetic field is set to be at an angle  $\theta_\epsilon=24\degree$ with the cavity wave vector $\vec{k}$.  By orienting the $\vec{B}$-field in this way we manage to engineer a system where   all three transitions between ground and excited $m_j$ levels have nonzero-coupling to the relevant cavity mode. The couplings  give rise to  single-photon Rabi frequencies  $2g_0 =2 g \sin \theta_\epsilon $ and $2g_\pm = 2g \cos \theta_\epsilon /\sqrt{2}$ for $m_j =0, \pm1$ respectively. 
Here $g = 2 \pi \times 3.5$~kHz.

For simplicity, we will consider the $i^\mathrm{th}$ atom as initially being in the ground state $\ket{\downarrow_i}\equiv \ket{g, p_i }$ with initial momentum $p_i$ along the cavity axis. The values of the $p_i$ reflect the finite temperature of the atoms and are drawn from a 1D Gaussian distribution with zero mean and rms momentum spread $\sigma_p$.  
To determine  $\sigma_p$ we perform time-of-flight expansion measurements after turning off the trapping lattice \cite{supp}. We find that the rms velocity spread along the cavity axis is $\sigma_v= 59(6)~\mathrm{mm/s}$ corresponding to a temperature of 38(8)~$\mu$K. 

We will first consider the case in which we use a laser at frequency $\omega_d$ that is resonant with $m_j=-1$ atomic transition frequency $\omega_{-}$ to resonantly excite this transition by driving the clockwise cavity mode, with $e^{i k Z}$ mode function (Fig.~\ref{fig1}d).
The drive is sufficiently large to establish an intracavity Rabi frequency $\Omega_d$ to nominally place each atom in a superposition of ground and excited states irrespective of initial momentum $p_i$.  The state after the nominal $\pi/2$ pulse is $\ket{\psi_{0,i}} = \tfrac{1}{\sqrt{2}}\left(\ket{\downarrow_i}+ \ket{\uparrow_i}\right)$, but with the excited state portion of the wavefunction $\ket{\uparrow_i}\equiv \ket{e, p_i+\hbar k }$ now displaced by the momentum of the absorbed photon $\hbar k$ due to the spin-orbit coupling \cite{lin2011spin, kolkowitz2017spin} during the excitation.  Here $\hbar$ is the reduced Planck constant and $k=2 \pi/\lambda$, see Fig.~\ref{fig1}f. This translates to the position space as imprinting a phase factor $e^{ikZ}$ on the atomic distribution.

We can  describe the $i^\mathrm{th}$ atom in terms of unitless single-particle pseudo-spin raising $\hat{s}_{i}^+=\ket{\uparrow_i}\bra{\downarrow_i}$ and  lowering $\hat{s}_{ i}^-= (\hat{s}^+_{ i})^\dagger$ operators, along with pseudo spin projection operators $\hat{s}_{i}^z = \tfrac{1}{2}\left(\ket{\uparrow_i}\bra{\uparrow_i}-\ket{\downarrow_i}\bra{\downarrow_i}\right)$, $\hat{s}_{x,i}= \tfrac{1}{2}\left(\hat{s}_{ i}^+ + \hat{s}_{ i}^-\right)$ and $\hat{s}_{i}^y= \tfrac{-i}{2}\left(\hat{s}_{i}^+ -\hat{s}_{ i}^-\right)$. Finally we define collective operators $\hat{S}_{\alpha}=\sum_{i=1}^N \hat{s}_{ i}^\alpha$ with  $\alpha\in \{x, y, z, +, - \}$. In the following, we will denote the expectation value of an operator with $A\equiv \langle\hat{A}\rangle$.

The coupling of the atoms to the cavity is described by the Tavis-Cummings Hamiltonian in the rotating frame of the atomic transition frequency $\omega_-$ as $\hat{H}_{TC}= \hbar g_{-} \left(\hat{S}_+\hat{a} + \hat{S}_-\hat{a}^\dagger\right) + \hbar \Delta_c^{-} \hat{a}^\dagger \hat{a}$.  The creation and annihilation operators $\hat{a}^\dagger$ and $\hat{a}$  describe the cavity mode that propagates in the same direction in which the initial atomic drive is applied. The counter-clockwise mode is neglected since the collective coupling to this mode is proportional to the expectation value of the collective operator $\frac{1}{N} \left\langle \sum_{i=1}^{N}  |\downarrow_i\rangle \langle e, p_i - \hbar k| \right\rangle$, which is zero for the initially prepared state and does not deviate from zero significantly due to collective or superradiant decay on the timescale of the experiment. The orthogonal s-polarized mode is detuned by 600~MHz due to mirror birefringence and thus can be safely neglected as well.

As experimentally demonstrated in \cite{norcia_2018_science_jkt, davis_2019_prl_mss}, in the large detuning limit $|\Delta_c^{-}|\gg\sqrt{N} 2 g_{-}$ \cite{muniz2020exploring}, the cavity mode can be adiabatically eliminated from the Tavis-Cummings Hamiltonian and one arrives at a cavity-mediated all-to-all spin-exchange Hamiltonian 

\begin{equation}
    \hat{H} = \hbar \chi_{-} \hat{S}_+\hat{S}_- + \hbar \sum_{i=1}^{N} \omega_i \hat{s}_{i}^z \, .
\label{eq:ExchangeHam}
\end{equation}

\noindent  The exchange interaction strength is set by the frequency $\chi_-= \frac{{g_{-}}^2}{\Delta_c^-}/\left(1+ \left(\kappa/\left( 2\Delta_c^{-}\right)\right)^2\right)\approx g_{-}^2/\Delta_c^-$ for the large detuning we work at $|2 \Delta_c^-/\kappa|>100$. 

The last term of Eq.~(\ref{eq:ExchangeHam} ) captures the Doppler broadening \cite{shankar2019squeezed} with  $\hbar \omega_i$ simply the kinetic energy difference between the ground and excited state portions of the wave function that is not common to all atoms $\hbar \omega_i = \hbar k p_i/m$, where $m$ is the mass of the atom.  The distribution of $\omega_i$ inherits the properties of the initial rms spread in momentum $\sigma_p$ such that the frequency distribution is described by a Gaussian distribution with zero mean and rms $\sigma_D = k \sigma_p/m \approx 2 \pi \times 87(9)$~kHz.  If one ignores interaction, the single-particle Bloch vectors $\vec{s}_i = \{s_{i}^x,s_{i}^y, {s}_{i}^z\}$ each precess at their own  frequency thus  shortening \ the norm of the total Bloch vector as a function of time with $|S_-(t)|^2 = |S_-(0)|^2 \,e^{-t^2/\tau_D^2}$ with $\tau_D = 1/\sigma_D = 1.8(2)~\mu$s, as illustrated in Fig.~\ref{fig1}e.

The all-to-all exchange interaction can be approximated as $\hat{S}_+\hat{S_-} \approx \hat{\vec{S}}\cdot\hat{\vec{S}} -\hat{S}_z^2$.  The first term means that there is an energy change associated with going to states of lower symmetry (Fig.~\ref{fig2}a), with a characteristic energy  gap $\hbar N \chi_-$ between adjacent states of total spin $S=N/2$ and $S=N/2 -1$.  This energy gap then competes with the single particle dephasing, protecting the spin alignment  when $N\chi_- \gtrsim \sigma_D$ by essentially pushing the single particle motional dephasing off-resonance.

\begin{figure}[t]
    \includegraphics[keepaspectratio, width=\columnwidth]{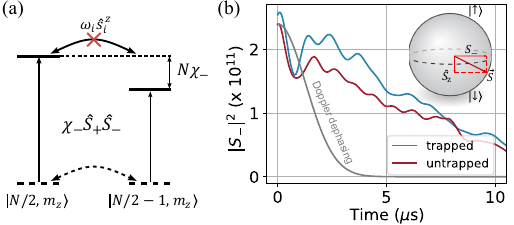}
    \caption{Many-body gap protection of squared atomic coherence $\left|S_-\right|^2$. (a) The cavity-mediated collective exchange interaction opens an energy gap $N\chi_-$ between two initially degenerate states (dashed lines) $\ket{S, m_z}$ of total spin $S$ and projection $m_z$.  The many-body energy gap pushes the single-particle motional dephasing (modeled by single-particle terms $\omega_i \hat{s}_i^z$) off-resonance such that the total $S$ remains constant and the atomic coherence is preserved.
    (b) The cavity field adiabatically follows the atomic coherence $S_-$ (see inset Bloch sphere) such that heterodyne detection of the weak leakage of field from the cavity allows us to determine $\left|S_-\right|^2$ versus time. After the 689~nm drive generates a $\pi/2$ pulse, Doppler dephasing would cause $\left|S_-\right|^2$ to decay rapidly as shown by the grey curve. The red curve is the experimentally observed coherence versus time when the atoms are untrapped.  The coherence extends well beyond the Doppler dephasing curve.  The extension of coherence is achieved by operating at a collective interaction scale $N\chi_-/2\pi= 430(10)$~kHz.  One also sees that the collective suppression of Doppler dephasing maintains coherence just as well as when one also imposes conventional single-particle trapping of the atoms in the Lamb-Dicke regime (blue curve).}  
    \label{fig2}
\end{figure}
To observe the suppression of Doppler dephasing, we load atoms into the intracavity optical lattice,  switch off the cooling lasers and optical lattice, and wait $10.1~\mu$s for the lattice light to decay such that the atoms are no longer trapped.  We note that radial expansion of the cloud and gravity can both be safely neglected for the approximate $40~\mu$s duration of the full measurement sequence.

We then apply a nominal $\pi/2$ drive pulse on the $m_j=-1$ transition. The cavity field adiabatically follows the collective optical dipole moment $a \approx S_- \left( g/\Delta_c\right)$.  As a result, we can infer $S_-$ by detecting the very small amount of light that leaks from the cavity using heterodyne detection.  We emphasize that $|S_-|/\left(N/2\right)$ indicates the contrast of a hypothetical Ramsey fringe if measured after phase evolution time $t$. We average between 100 to 1200 repetitions of the experiment with more averaging when using lower atom numbers. 

In Fig.~\ref{fig2}b, the grey curve shows the predicted decay of $|S_-|^2$ for Doppler broadening as determined from the time of flight measurements without interaction.  The red data curve shows that the measured coherence extends well beyond this timescale, directly indicating that Doppler dephasing is being suppressed by the interaction.

For comparison, we repeat the same experiment as above but leave the optical lattice on such that the tight confinement of the lattice places the atoms in the traditional Lamb-Dicke regime suppressing Doppler dephasing via single-particle physics.  The blue data curve of Fig.~\ref{fig2}b for this case closely resembles the original Doppler-sensitive red data curve where in contrast, the insensitivity arises from the collective interactions mediated by the cavity.  Both curves decay more quickly than the excited state single-particle decay time of $\tau=1/\gamma=21~\mu$s, indicating that the excess decay primarily appears to arise from something other than Doppler dephasing.

\begin{figure}
    \includegraphics[keepaspectratio, width=\columnwidth]{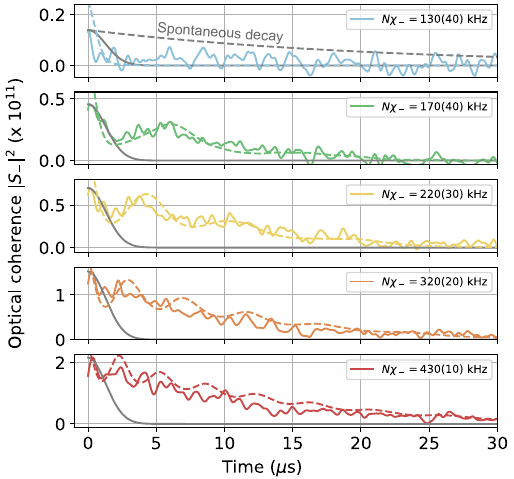}
    \caption{Emergence of collective gap protection against Doppler dephasing by varying $N\chi_-$ via adjusting the number of atoms loaded into the cavity. With a small $N\chi_-$, coherence collapses more rapidly than the spontaneous decay (grey dashed curve), but increasing the atom number provides full protection against Doppler dephasing. Faster-than-expected coherence decay at small $N\chi_-$ is attributed to an additional spatial magnetic field inhomogeneity. The colored dashed curves are the numerical simulations using a mean-field simulation that includes interactions, dephasing, spontaneous decay, and collective decay.} \label{fig3}
\end{figure}

For the Doppler dephasing to be suppressed by the exchange interaction, the charateristic gap frequency should be larger than the dephasing frequency scale $N \chi_- > \sigma_D$. We observe this behavior in the data of Fig.~\ref{fig3}, where we show the measured $|S_-|^2$ versus time for a range of $N\chi_-/2\pi$ from 130 to 430~kHz. We determine $N\chi_- $ by measuring dispersive shifts of the cavity-like mode at large $N$ and then using this measurement to calibrate a fluorescence detection for smaller $N$ \cite{supp}.  The gap frequency is adjusted by varying the number of atoms at the initial 2D blue MOT stage to avoid changing the temperature of the atomic ensemble. The minimum number of atom is limited by signal to noise in measuring $|S_-|^2$ since the detected signal scales as $N^2$.

For the smallest value of $N\chi_-/2\pi$ = 130(40)~kHz, $|S_-|^2$ rapidly collapses with 1/e time 0.9(3)\us. However, one can see that as $N\chi_-$ increases slightly, the coherence begins to last substantially longer, with $|S_-|^2$ reaching a $1/e$ value at $10~\mu$s at the largest value of  $N\chi_-/2\pi$ = 430(10)~kHz.  When the frequency gap scale approaches the total dephasing frequency (see below) at $N\chi_-/2 \pi$ = 170 to 270~kHz, an oscillation with a frequency comparable to $\chi_-N$ emerges in $|S_-|^2$.  This oscillation can be identified as a damped Higgs-like oscillation \cite{young_2024_nature_jkt,luo_momentum-exchange_2024}.

The observed decay of $|S_-|^2$ at $N\chi_-/2\pi = 130$~kHz is more rapid than that predicted for Doppler dephasing (grey curves). We believe that the Doppler dephasing prediction is accurate, and instead assign  this to a magnetic field gradient across the atomic cloud that leads to a dephasing with rms frequency $\sigma_B = 2\pi\times 157$~kHz. 
Note that the exchange interaction also suppresses this dephasing as has been observed previously \cite{Scherg2020,young_2024_nature_jkt}.  

The dashed numerical simulation curves in Fig.~\ref{fig3}  agree well with the observations.  The simulations use a mean field master equation that includes the Hamiltonian dynamics of Eq.~\ref{eq:ExchangeHam}, single-particle spontaneous emission described by  jump operators $\sqrt{\gamma/2}\hat{s}_{i}^-$, and  collective decay  described by a jump operator $\sqrt{\Gamma_s/2} \hat{S}_-$, with  $\Gamma_s\approx \kappa \left(g_{-}/\Delta_c^{-}\right)^2$.  For the simulations, the atom number $N$ was adjusted to better describe each curve, with no adjustments at the largest atom numbers and an adjustment of $N$ downwards by 40\% at the smallest atom numbers.  We believe that this small discrepancy may arise from an inaccuracy in the transfer of the  calibration of $N \chi_-$ performed at the highest atom number but extended to lower atom number via fluourescence imaging of the atoms.

\begin{figure}
    \includegraphics[width=\columnwidth]{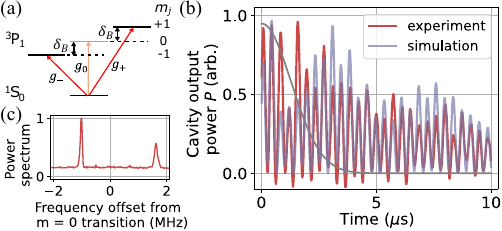}
    \caption{Many-body gap protection and beating of the coherence with several excited states. (a) We simultaneously excite $m_j=\pm1$ with Zeeman splitting $\delta_B$ by setting the drive to $\omega_d=\omega_0$ and increasing the drive's Rabi frequency. 
    (b) The power emitted from the cavity, expressed as a total optical atomic coherence $|S_-|^2$ versus time shows beating at $2 \delta_B$, persisting well beyond the grey Doppler dephasing curve, demonstrating that the coherence on each transition is gap-protected against Doppler dephasing. The gap protection for this data was set to $ N\chi_- = 430(10) \kHz$.
    (c) The power spectrum of the detected light from t = 0 to 10~$\us$ shows two discrete Fourier frequencies indicating that the two Zeeman transitions independently radiate light.
  }
    \label{fig4}
\end{figure}

We can also observe gap protection against Doppler dephasing when we simultaneously excite primarily $m_j=\pm1$ (Fig.~\ref{fig4}a).  In Fig.~\ref{fig4}b, we show the measured cavity output power versus time. We see rapid oscillations with an envelope persisting well beyond the timescale for single-particle Doppler dephasing. Exchange interactions lock each of the $m_j=\pm1$ ensembles individually but not strongly enough to bind them together. This gives rise to beating in  the output power at $2\delta_B$. We have seen similar oscillations in a standing wave cavity with trapped atoms, emulating phase phases of BCS superconductors \cite{young_2024_nature_jkt,young2024timeresolved}. Here, both ensembles share the same ground state, and the oscillations are more pronounced and last longer due to the homogeneous cavity coupling. The power spectrum of the light (Fig.~\ref{fig4}c) shows light is being radiated largely independently from the two $m_j=\pm 1$ levels.

We have experimentally demonstrated that cavity-mediated interactions can suppress Doppler dephasing on a narrow optical clock transition at 400~THz.
Beyond being of fundamental interest, this provides a potentially new tool for optical metrology and spectroscopy \cite{supp}, such as searches for new particles and fields through the modulation of optical atomic transition frequencies \cite{SafranovaSearch2018}. The approach here also simultaneously enables measurements at timescales fast compared to the phase evolution time.

In the future, exploring quantum simulation in this system with homogeneous atom-cavity coupling will also be of great interest, avoiding complications due to inhomogeneous coupling in standing wave cavities. The two directionally indepedent  cavity modes also provides a controllable degree of freedom that bares some similarity to tunable mode changing collisions in quantum many-body systems.
It will also be interesting to connect to models for neutrino-neutrino interactions that give rise to a multitude of collective effects in flavor space \cite{balantekin2023quantum}. By using three atomic  levels to emulate three neutrino flavors, and  both clockwise and counter-clockwise cavity modes to controllably emulate trajectory-dependent interactions, we should be able to recreate  simplified versions  of such a model system.

\begin{acknowledgments}

This material is based upon work supported by the U.S. Department of Energy, Office of Science, National Quantum Information Science Research Centers, Quantum Systems Accelerator. We acknowledge additional funding support from the National Science Foundation under Grant Nos. 2317149 (Physics Frontier Center) and OMA-2016244 (Quantum Leap Challenge Institutes), the Vannevar Bush Faculty Fellowship, the  Heising-Simons Foundation, and NIST.
\end{acknowledgments}

\bibliographystyle{apsrev4-2}
\bibliography{bib}

\clearpage

\title{Supplementary Information: Many-body gap protection of motional dephasing of an optical clock transition}

\maketitle
\onecolumngrid

\renewcommand{\thefigure}{S\arabic{figure}}
\setcounter{figure}{0}
\renewcommand{\theequation}{S\arabic{equation}}
\setcounter{equation}{0}
\setcounter{section}{0}
\setcounter{page}{1}
\setcounter{secnumdepth}{2}

\section{Time-of-flight measurements}

To measure the $rms$ axial velocity of the atoms, which determines the predicted Doppler dephasing time scale, we perform time-of-flight measurements.  The measurements begin by shutting off all laser cooling beams and allowing any untrapped atoms to fall away under gravity for 40~ms. We then the turn off the 813~nm lattice beams in less than 200~ns.  The cavity power at this wavelength then exponentially decays with a characteristic decay time of $\tau_L= 240$~ns. The lattice depth reaches less than $1~\mu$K in $1.2~\mu$s, negligible to the subsequent time of flight periods spanning out to 20~ms.

The cloud is allowed to ballistically expands along the cavity axis for a variable time $t_f$ spanning 0 to 20~ms.  We then use resonant 461 nm light to capture a fluorescence image of the atomic cloud which we fit to determine the rms cloud size $\sigma_z (t_f)$  size along the cavity-axis, as well as the center of the atomic cloud versus time.  The falling of the center of the cloud with gravity is used to calibrate the image dimensions. The measured $\sigma_z(t_f)$ versus $t_f$ is well described by the fit function $\sigma_z(t_f) = \sqrt{\sigma_0^2 + (\sigma_v t_f)^2}$, with the fitted rms axial velocity $\sigma_v$ reported in the main text, and $\sigma_0$ capturing the initial rms cloud size along the cavity axis.

\section{Calibration of $\chi_- N$}
To characterize the exchange interaction strength $N\chi_-$,  we exploit the fact that the full Hamiltonian of the system includes a previously neglected term $\hat{H_c}=\hbar \left[\Delta_c+ \chi_- \left(\hat{N}_\downarrow -\hat{N}_- \right) + \chi_0 \left(\hat{N}_\downarrow -\hat{N}_0 \right) \allowbreak + \chi_+ \left(\hat{N}_\downarrow -\hat{N}_+ \right)  \right]\hat{a}^\dagger \hat{a}$ where the operators $\hat{N}_\alpha$ are projection operators that count the number of atoms in a given atomic state.  One sees that this Hamiltonian couples the inversion on each transition to the resonance frequency of the cavity.  When all of the atoms are in the ground state, this results in a shift of the cavity frequency $\delta_c= \omega_c' -\omega_c = N \left(\chi_-+\chi_0+\chi_+\right)$ from which we can solve for $N\chi_-$ since we know the coupling strengths and detunings. The cavity frequency shift $\delta_c$ is measured by applying a short pulse of drive light at the bare cavity resonance frequency and then detecting the subsequent ring down of the cavity field with the heterodyne detection.  This is repeated with and without atoms to determine $\delta_c$, and varying the drive light amplitude to ensure that the atoms are not significantly being excited out of the ground state.  This calibration is performed at $\Delta_c = - 2\pi \times 5$~MHz and at the highest atom number to in turn calibrate a fluourescence imaging system to scale to lower atom number.

\section{Scaling of gap protection}

We consider the scaling of gap protection against single particle dephasing. First, we will  consider a cavity with infinite finesse $F\rightarrow\infty$ such that $\kappa\rightarrow 0$ and collective decay $\Gamma_s\rightarrow 0$.  We also assume a simple two-level model.  To be in a regime in which the cavity can be adiabatically eliminated, we require that the cavity detuning is larger than the collective vacuum Rabi splitting, so we take $\Delta_m= \alpha_V \sqrt{N} 2 g$, where $\alpha_V\gg 1$ is a number of order 4 to 10. This minimum allowed detuning maximizes the interaction energy scale $\chi N$.  For gap protection against motional dephasing, we want $\chi N =\alpha_G \sigma_D$ with  $\alpha_G\sim 2$, i.e.~the interaction scale at least two times larger than the rms dephasing rate $\sigma_D$.  The maximum dephasing that can be suppressed is $\sigma_{max}= \tfrac{\sqrt{N} g}{2\alpha_V\alpha_G}$. For reference, neglecting collective and single particle decay, the infinite time coherence is numerically simulated and found to be $\tfrac{2}{N}|S_-| = 0.5, 0.79$ and $0.90$ at $\alpha_G = 1.25, 2$, and $3$ respectively.

We now ask if we turn down the finesse $F$ to a finite value, when does the collective superradiant decay rate per particle for a Bloch vector at the equator equal the rate of dephasing $\Gamma_s = \sigma_{max}$, giving a minimum required finesse $F_{min} = \tfrac{\alpha_G}{8\alpha_V}\tfrac{\omega_{FSR}}{\sqrt{N} g}$. The free spectral range of the cavity is $\omega_{FSR}= 2 \pi c/ L$, where for a ring cavity $L$ is the round trip length.  This is the minimum cavity finesse needed to start to improve the coherence time relative to the original dephasing at rate $\sigma_{max}$.  One can also ask what is the maximum useful finesse by considering when the collective decay per particle is equal to the single-particle spontaneous emission $\Gamma_s= \gamma$.  One finds that spontaneous emission sets a maximum $F_{max} = \omega_{FSR}/4 \gamma \alpha_V^2$, above which there is limited value in further increasing the finesse though there is no fundamental harm in doing so.

If the actual dephasing rate in the system is given by $\sigma_D\le \sigma_{max}$, then one should operate at a larger cavity detuning $\Delta_D =\Delta_{m} \left(\sigma_{max}/\sigma_{D}\right)$ and one finds minimum and maximum finesses $F_{min, D}=F_{min}\left(\sigma_D/\sigma_{max}\right)$ and  $F_{max, D}=F_{max}\left(\sigma_D/\sigma_{max}\right)^2$.  The ratio of the actual finesse to the minimum finesse $F/F_{min,D}$ gives the ratio of the superradiance-limited coherence time relative to the original decoherence time scale $1/\sigma_D$. 

We wish to concretely connect the above abstract Hamiltonian parameters to actual experimental parameters. For a two level system on axis of a ring cavity, the coupling is given by $g^2 = \tfrac{3}{8 \pi}\tfrac{\lambda^2 c}{V_c} \gamma$, where the transition wavelength is $\lambda$, the cavity mode volume is $V_c = \pi w_0^2 L/2$, and $w_0$ is the smallest cavity mode waist (i.e. $1/e^2$ in intensity.)  

In the experimental system here, the calculated maximum dephasing rate that can be enhanced is $\sigma_{max}/2 \pi = 180$~kHz or a temperature of 170~$\mu$K.  The minimum and maximum finesses are $F_{min}= 43$ and $F_{max} = 4.2\times 10^3$, to be compared to the actual finesse $F= 6.2\times 10^3$.  The above assumes  $N = 10^6$, $\alpha_G=2$ and $\alpha_V = 4$.  Correcting for the geometric modification ($g_-=g \cos\theta_p/\sqrt{2}$) decreases $\sigma_{max}/2\pi$ to 120~kHz and increases $F_{min}$ to 66 with a cavity detuning $\Delta_{m}/2\pi = 15$~MHz.

As an example for an optical transition narrower than the one utilized here, we consider the following:  an atomic transition with $\gamma/2\pi = 10$~Hz, $\lambda= 700~$nm, mass of $88$~u, a ring cavity with $L=10$~cm, $w_0 = 100~\mu$m,   $N = 10\times 10^6$, $\alpha_G=2$ and $\alpha_V = 4$.  One finds $\sigma_{max}/2 \pi = 21$~kHz which if the dephasing is due to Doppler broadening is a temperature of $2.3~\mu$K. The minimum finesse is $F_{min,D} =3.9\times 10^2\left(\sigma_D/\sigma_{max}\right)$ and the maximum finesse is $F_{max, D} = 3.1\times 10^6 \left(\sigma_D/\sigma_{max}\right)^2$.

\end{document}